\documentclass[showpacs,twocolumn]{revtex4}
\usepackage{graphicx}

\newcommand{\rhoc}{\rho_{\mbox{\small cp}}}

\begin{document}
\title{
A rejection-free Monte Carlo method for the hard-disk system
}

\author{Hiroshi Watanabe$^{1,2}$\footnote{E-mail: 
hwatanabe@is.nagoya-u.ac.jp}, 
Satoshi Yukawa$^2$\footnote{%
Present address,
Department of Earth and Space Science, Graduate School of Science,  
Osaka University
},
M. A. Novotny$^3$
and Nobuyasu Ito$^2$}
\affiliation{$^1$ Department of Complex
Systems Science, Graduate School of Information Science,
Nagoya University, Furouchou, Chikusa-ku, Nagoya 464-8601, Japan}
\affiliation{$^2$ Department of Applied Physics, School of Engineering,
The University of Tokyo, Hongo, Bunkyo-ku, Tokyo 113-8656, Japan}
\affiliation{$^3$ Department of Physics \& Astronomy, 
and HPC$^2$ Center for Computational Sciences, 
Mississippi State University, Mississippi State, Mississippi 39762-5167, USA}

\begin{abstract}
We construct a rejection-free Monte Carlo method 
for the hard-disk system. 
Rejection-free Monte Carlo methods preserve the 
time-evolution behavior of the standard Monte Carlo method,
and this relationship is confirmed for our method
by observing nonequilibrium 
relaxation of a bond-orientational order parameter.
The rejection-free method
gives a greater computational efficiency 
than the standard method at high densities.  
The rejection free method is implemented in a shrewd manner using 
optimization methods to calculate a rejection probability 
and to update the system.
This method should allow an efficient study of the dynamics of 
two-dimensional solids at high density.
\end{abstract}


\maketitle

\section{Introduction}

Monte Carlo (MC) methods have become more powerful tools with 
the development of faster and more accessible computers.  
Many different phenomena have been studied with 
MC methods \cite{Binder,Landau}.  The melting behavior of the hard-disk 
system \cite{Alder} is one of such subjects
which has been studied with Monte Carlo methods
\cite{LeeStrandburg, Sengupta, Fernandez, Zollweg, Jaster, Weber}.
Consider the hard-disk system with $N$ particles.
The Monte Carlo method for the hard-disk system has the following steps.
First, choose one particle randomly, {\it i.e.} choose it
with a probability of $1/N$.
Then choose a new position for the center of the chosen particle, the 
new position is chosen uniformly in the circle with radius $\sigma_s$ 
centered on the original position of the particle.  
The trial move is accepted when the new position has no overlap with 
other particles, otherwise, the trial move is rejected.

The MC methods introduced above have been used to obtain the equilibrium
state of the system. 
To study the equilibrium state of the system, the value of $\sigma_s$, 
the trial step length, is often chosen in order that the rejection rate 
is near 50\%.  
Recently, Watanabe {\it et al}. \cite{Watanabe,Watanabe2}
studied the hard-disk system by observing nonequilibrium
behavior using a molecular dynamics (MD) method.
In order to compare the results of MD and MC,
we have to consider the dynamics of the MC method.
The dynamics of the MC method for the hard-disk system 
can be understood as a Brownian motion, like pollen particles in a 
fluid.  
The value of $\sigma_s$ corresponds to the amplitude of the external 
random force. Therefore, it determines a time scale of this system.
If values of $\sigma_s$ are not identical for all densities, 
we cannot compare the dynamics, like relaxation phenomena and fluctuations, 
between different densities.
However, keeping $\sigma_s$ fixed leads to high rejection rates 
in the Monte Carlo at high densities.  
A similar problem occurs for spin systems
both at low temperatures and in a strong external field:  
a rejection rate becomes very high, so a huge number of 
trials is required to make a change.
This problem can be avoided using a different 
technique called the rejection-free Monte Carlo (RFMC) method.
The RFMC method was first constructed for discrete spin systems \cite{Bortz},
for a review see \cite{Novotny}. 
Miyashita and Takano \cite{MiyashitaTakano} applied the RFMC 
method to the kinetic Ising model
in order to study dynamical critical behavior.
Recently, Mu{\~n}oz {\it et al.}~\cite{Munoz} proposed 
a new RFMC algorithm which can treat models with 
continuous degrees of freedom.
In this paper, we construct and utilize a RFMC algorithm for the 
hard-disk system
based on the method of Ref.~\cite{Munoz}.

\section{Methodology}

\subsection{A Rejection-free Monte Carlo method}

\begin{figure}[htb]
\begin{center}
\includegraphics[width=.7\linewidth]{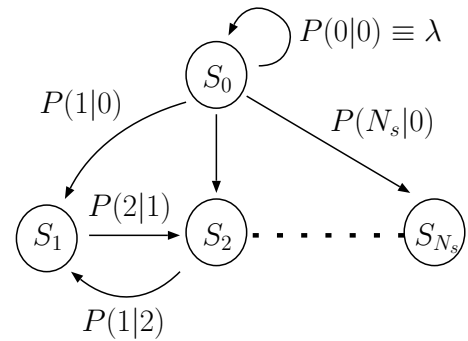}
\end{center}
\caption{
A Markov chain of Monte Carlo steps.
}
\label{fig_markov}
\end{figure}

A Monte Carlo method is an implementation of a Markov 
process on a computer, and hence is sometimes called 
a Markov Chain Monte Carlo.  
The Monte Carlo method calculates various physical quantities by 
updating states of a system using random variables.
These updating processes can be illustrated by a 
Markov chain (see the schematic in Fig.~\ref{fig_markov}).
Let the current state be at $S_0$ and 
the states possible to move from $S_0$ are denoted by 
$S_i (i=1,2,\cdots,N_s)$.
Define $E_i$ as the energy of the state $S_i$.
The new state $S_i (i=0,1,\cdots,N_s)$ will be chosen
with probability $P(i|0)$. 
One way to ensure that the system will relax to the 
equilibrium state is to insist that the probability $P(i|0)$ 
satisfy the detailed balance condition \cite{Landau}.  
 
One of the well-known ways \cite{Landau} to satisfy the 
detailed balance condition is to use a heat-bath transition probability.
In the heat-bath method, the probability $P(i|0)$ is defined to be
\begin{equation}
P(i|0) = \frac{\exp{(-\beta E_i)}}{\sum_{k=0}^{N_s} \exp{(-\beta E_k)}}.  
\label{eq_heatbath}
\end{equation}.
When a system has a continuous degree of freedom,
the summation of Eq.~(\ref{eq_heatbath}) becomes an integration
which is generally difficult to calculate analytically.

Another popular way to satisfy the detailed balance condition 
is a Metropolis method.
In this method, each step contains
two parts; selecting a new state and accepting or rejecting
the trial to move to the selected state.
First, pick up a state $S_i$ from all possible states to move to
with uniform probability $1/N_s$. 
The probability $P(i|0)$ to move from $S_0$ to $S_i$ is 
defined to be $1$ when $\Delta E_i < 0$, otherwise it is 
$\exp{\left(-\beta \Delta E_i \right)}$ with the energy difference 
$\Delta E_i \equiv E_i -  E_0$.
Therefore, the probability $P(i|0)$ in the Metropolis method is 
\begin{equation}
P(i|0) = 
\left\{
\begin{array}{cc}
1/N_s & \mbox{if} \quad \Delta E_i \le 0, \\
\exp{\left(-\beta \Delta E_i \right)}/N_s  & \mbox{otherwise}.
\end{array}
\right.
\end{equation}
When a trial is rejected, the configuration of the system 
is not updated.
The probability $\lambda \equiv P(0|0)$ to stay in the current state 
after the trial
is given by the expression 
\begin{equation}
\lambda = 1 - \sum_{i\neq 0} P(i|0).
\label{DefLambda}
\end{equation}
For some parameters, {\it e.g.} under a strong external field and 
at an extremely low temperature, 
the value of $\lambda$ can be very nearly $1$.
In such cases, most of computational time is spent on calculating 
trials which will be rejected. This rejection rate drastically 
decreases the efficiency of 
the computation.

In order to overcome this problem, 
a rejection-free Monte Carlo (RFMC) method is proposed.
It is an example of an event driven 
algorithm~\cite{Binder,Landau} and has also been 
called a waiting time method \cite{MiyashitaTakano,Jesper}.
Each steps of the RFMC method involves first computing the time to leave the 
current state (the waiting time $t_{\mbox{wait}}$),
and then choosing a new state to move to with the appropriate probability.
It does not contain the judgment to accept or reject a trial,
and, therefore, it achieves rejectionless updates of the system in each 
algorithmic step.
The waiting time $t_{\mbox{wait}}$ is calculated using a (pseudo) 
random variable.
The probability $p(t)$ to keep staying at the current state for $t$ steps 
decays exponentially,
\begin{equation}
p(t) = \lambda^t
     = \exp{(t \ln{\lambda})},
\end{equation}
with $\lambda$ defined in Eq.~(\ref{DefLambda}).
Note that $\ln \lambda <0$ since $0<\lambda<1$.
Inversely, the time $t$ to stay 
in the current state is determined
with a uniform random number $r$ on $(0,1)$ to be
\begin{equation}
t_{\mbox{wait}} = \left\lfloor \frac{\ln r}{\ln \lambda} \right\rfloor +1,
\label{DefTwait}
\end{equation}
where $\lfloor x \rfloor$ denotes the integer part of $x$ and
the rounding down is introduced to express the discrete time step 
in MC \cite{Novotny,Munoz}.

After the time of the system is advanced by $t_{\mbox{wait}}$, 
a new state $S_i$ is chosen from the all states possible to move to
except for the current state with the probability proportional to 
$P(i|0)$ \cite{NovotnyPRL,Novotny,Munoz}.
Since all values of $P(i|0)$ are required to proceed one algorithm step
in the RFMC, the computational cost of one step is higher than 
the normal MC. 
However, the waiting time $t_{\mbox{wait}}$, the time which 
can be advanced in one algorithmic step,
becomes large at low temperatures, and consequently 
the efficiency of the RFMC 
can become better than that of standard MC.

\subsection{Application to hard-disk systems}

\begin{figure}[htbp]
\includegraphics[width=.8\linewidth]{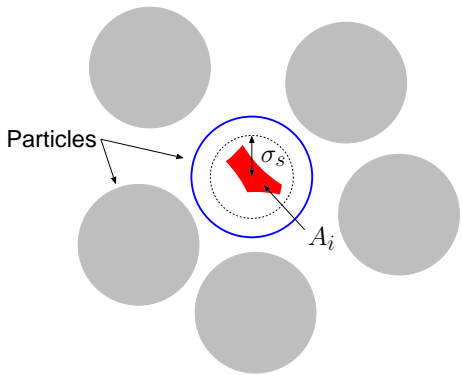}
\caption{
(Color online) A schematic drawing of the definition of $A_i$ (shaded).
The solid circles are particles and the small dashed circle has 
a radius $\sigma_s$. The shaded area is 
the area which is a continuous set of the points 
that the center of the chosen particle can move to.
The ratio of $A_i$ to the area of 
the trial circle $\pi {\sigma_s}^2$
gives the probability of accepting the move, $1-\lambda_i$, given 
that the center particle has been chosen as the one to move.
}
\label{fig_ai}
\end{figure}

Consider a hard disk system with $N$ particles with the radius $\sigma$.
A standard MC method for the system involves
choosing a particle, and trying to move the chosen particle
within a circle with radius $\sigma_s$ centered on the 
original position of the chosen particle.
To apply a RFMC method to the hard-disk system,
define $\lambda_i$ as the probability that 
a trial to move particle $i$ is rejected (given that particle $i$ was 
chosen as the particle to attempt a move).
Using the definition of $\lambda_i$, we can construct 
the algorithm of the RFMC method for the hard-disk system as follows:
\begin{enumerate}
\item Calculate the waiting time $t_{\mbox{wait}}$
using Eq.~(\ref{DefTwait}) with $\lambda = \frac{1}{N} \sum_i \lambda_i$.
\item Advance the time of the system by $t_{\mbox{wait}}$.
\item Choose a particle $i$ with the probability proportional
to $1-\lambda_i$, which is the probability that 
(given that particle 
$i$ was the particle chosen for an attempted move) 
the trial to move the particle $i$ would be accepted.
\item Choose the new position of the chosen particle $i$
uniformly from all the points to which the particle $i$ is allowed to move.
\end{enumerate}
The steps described above are the same as the RFMC for 
continuous spin systems \cite{Munoz}, 
but the algorithms to calculate $\lambda_i$ 
and to determine a new position of the chosen particle are unique to 
the hard-disk system.
The probability $\lambda_i$ can be calculated to be,
\begin{equation}
\lambda_i = 1 - \frac{A_i}{\pi {\sigma_s}^2},
\label{DefLamda_i}
\end{equation}
with an area $A_i$ into which the particle $i$ is 
allow to move (see Fig.~\ref{fig_ai}).
Therefore, if we can calculate $A_i$ for all particles, then
we can construct the RFMC algorithm for the hard-disk system.

\subsection{Calculation of $A_i$}
\label{sec_calcarea}

The area $A_i$ is the continuous set of positions
in which the particle can be placed without any overlaps.
Without neighboring particles, the shape of $A_i$ would be a filled circle
with a radius $\sigma_s$. Let's call it a trial circle.
In the general case, the shape of $A_i$ is the remaining part
of the trial circle after removing the overlap of \lq shadows'
of neighboring particles.  
The shape of the shadow is a circle with a radius $2\sigma$ 
which is concentric to a neighboring particle.
Let's call this a shadow circle.
The area $A_i$, thus, consists of areas of arcs of a trial circle 
and that of shadow circles.  

\begin{figure}[htbp]
\noindent
\includegraphics[width=0.45\linewidth]{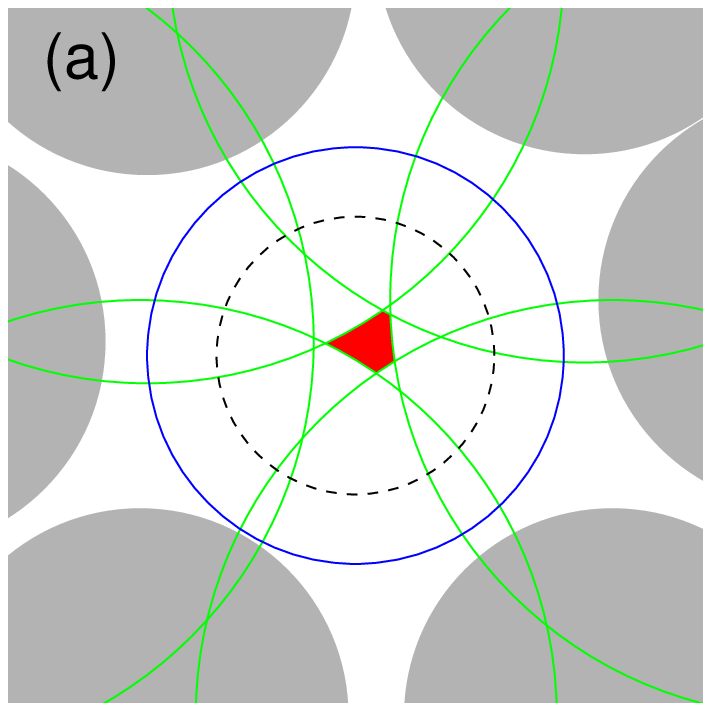}
\includegraphics[width=0.45\linewidth]{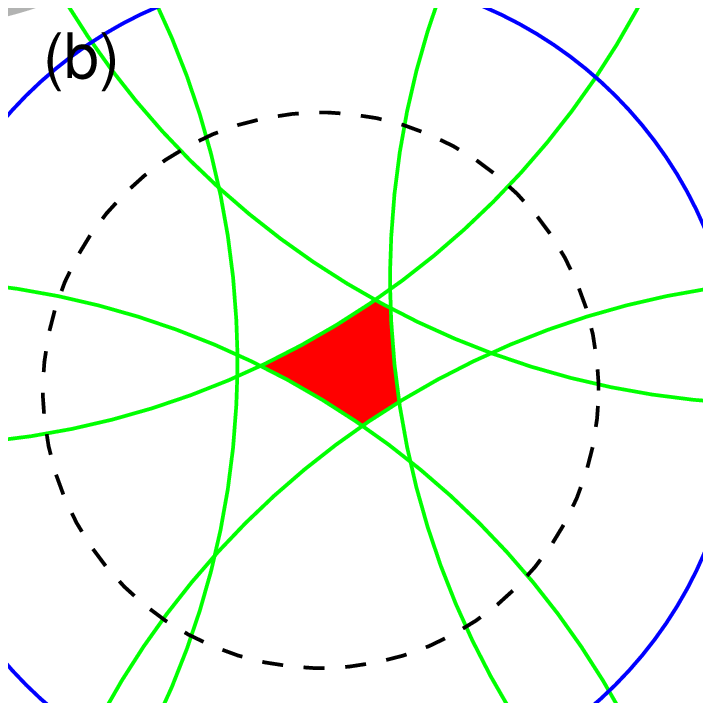}
\noindent
\includegraphics[width=0.45\linewidth]{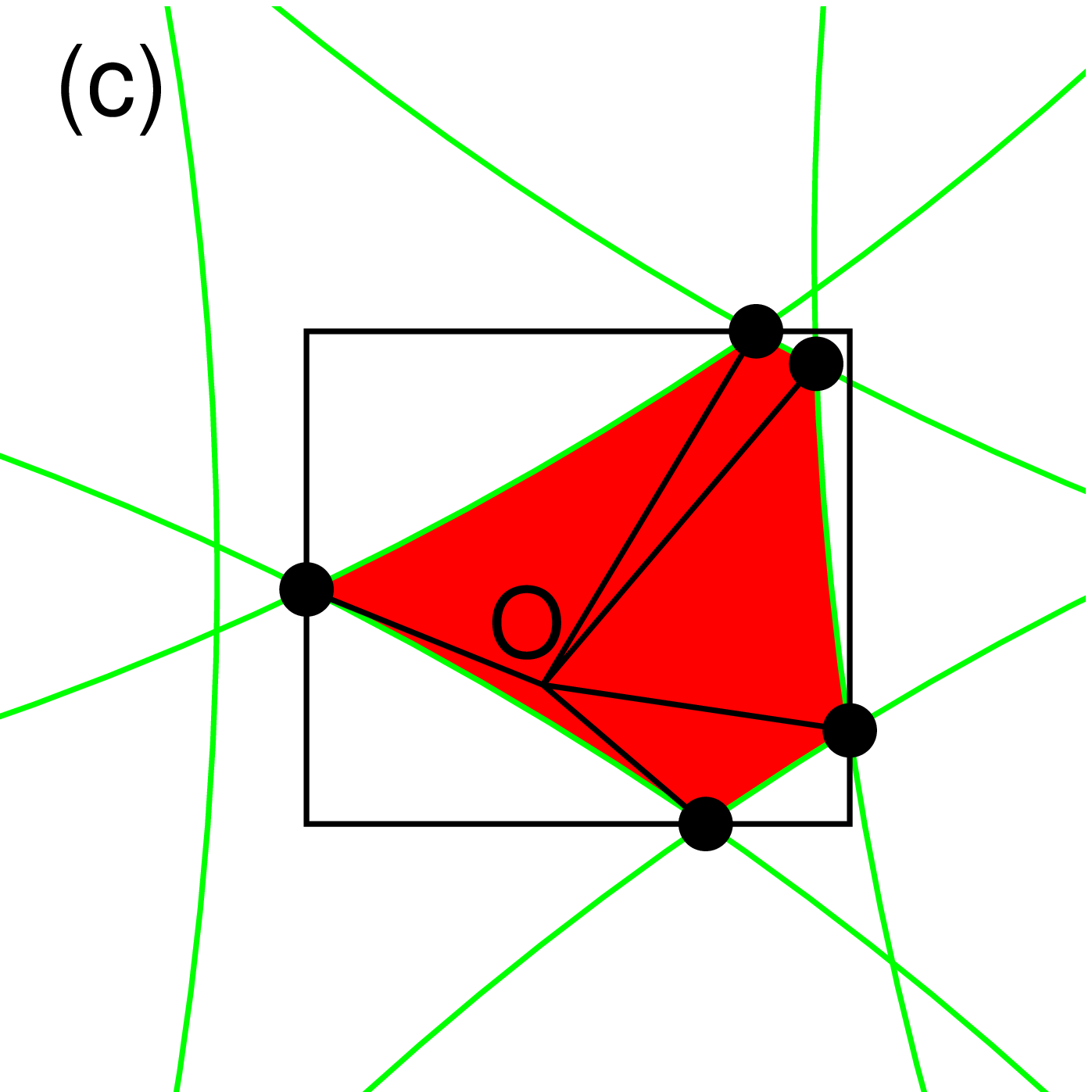}
\includegraphics[width=0.45\linewidth]{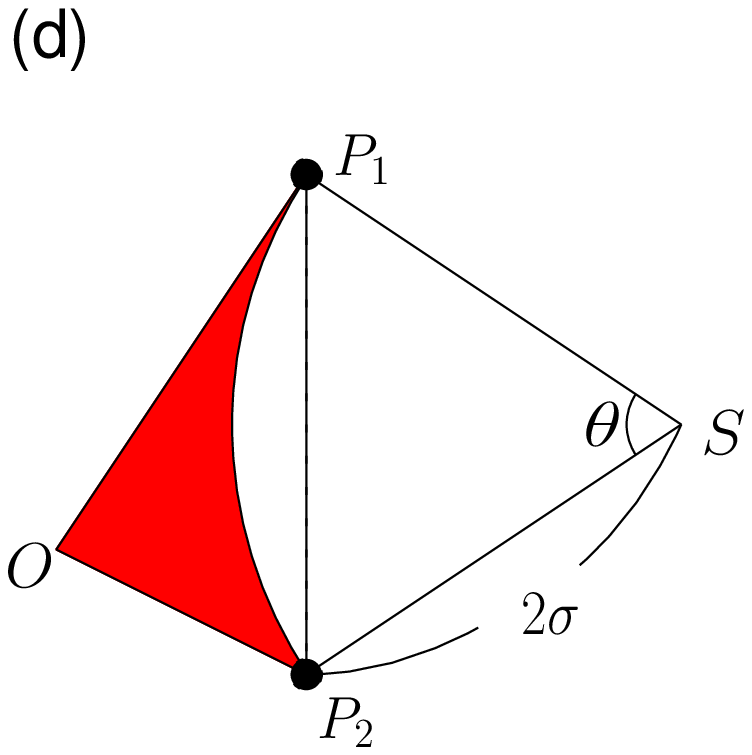}
\caption{
(Color online) Computation of the value of $A_i$.
(a) Filled gray circles represent neighboring particles with radius $\sigma$
and large circles are shadows of them (we call them \lq shadow circles') 
with radius $2 \sigma$.
(b) An enlargement. The shaded area is the area
into which the particle $i$ is allowed to move.
This figure is made by removing overlaps of shadow circles from 
the trial circle of radius $\sigma_s$ centered on the chosen particle.
(c) The survived vertex points. 
The center of the trial circle is denoted by $O$. 
The solid circles represent survived vertex points, 
which form the area $A_i$. With them, we can calculate
the value of $A_i$. The rectangle denotes a bounding rectangle.
Each two adjacent survived vertex points and the center point O form 
a triangle.  
In this example there are five survived vertex points, and consequently five 
triangles to consider.  
(d) To calculate a portion of $A_i$, the area within 
each triangle formed by survived vertex points and O is calculated.  
The survived vertex points are the intersection points of the 
shadow circles or the trial circles, and here 
are denoted by $P_1$ and $P_2$. The center of the shadow particle is $S$.
To find the shaded area a Monte Carlo procedure is performed in the 
shaded area of either (c) or (d), and only survived points generated in 
the shaded area are used as the new location for the new point O.  
}
\label{fig_survival}
\end{figure}

To compute the value of $A_i$, we develop a method we call the 
survival point method.
See Fig.~\ref{fig_survival}~(a). 
The chosen particle is shown as a solid circle, the trial circle is 
shown as a concentric dashed circle, and the area $A_i$ 
is the shaded region.  
Each neighboring particle (filled circles) has a 
shadow circle which is concentric and has radius $2\sigma$.  
An enlargement of the area $A_i$ is shown in Fig.~\ref{fig_survival}~(b).
It is seen that in this example this area has five vertices which 
are intersection points of shadow particles, we call them survived 
vertex points.
In Fig.~\ref{fig_survival}~(c), these survived vertex points are shown
as small filled circles.  Straight lines connect the center of 
the chosen particle and the intersection points. In this case the 
area $A_i$ is divided into five figures. 

Each divided figure is the remaining part of an isosceles triangle with 
the overlap of a shadow circle removed.  It is easy 
to calculate this area. 
Thus, all we have to do is to find all survived vertex points which 
form the area $A_i$.
First, make a list of all intersection points of shadow circles 
and the trial circle.  Next, remove points which are included in 
other shadow circles from the list, since these points cannot be vertices
forming the area $A_i$.
After this removal process, we have the vertices which form 
the area $A_i$ (see Fig.~\ref{fig_survival}~(c)).
The calculation process of a partial figure which forms $A_i$
is shown in Fig.~\ref{fig_survival}~(d).
The vertices are denoted by $P_1$ and $P_2$, and
the center of the shadow circle is denoted by $S$.
The survived vertices $P_1$ and $P_2$ are on the shadow circle centered at 
$S$, so $\overline{SP_1} = \overline{SP_2} = 2\sigma$. 
The area of $O P_1 S P_2 $ can be calculated by summing the 
two triangles $O P_1 P_2$ and $S P_1 P_2$ with Heron's formula.
The area of the chord is $4 \sigma^2 \theta$.
Finally the portion of the area 
$O {P_1}{P_2}$ is calculated by subtracting the area of the chord
$S {P_1} {P_2}$ from the area of the quadrilateral $O P_1 S P_2$.
The total area $A_i$ is the sum of one such calculation for each 
survived vertex.

\subsection{Choosing a particle to move}

After calculation of $t_{\mbox{wait}}$ and advancing the 
time of the system by it,
we have to choose a particle $i$ to move 
with a probability proportional to $1-\lambda_i$.
With a direct implementation, {\it i.e.}, with the integration 
scheme~\cite{IntegrationSearch},
the order of the computation is $O(N)$,
which is very time consuming.
Other approaches are proposed like a three level 
search for spin systems \cite{threelevel}.
The three-level search improves the efficiency of the search 
by determining coordinates of a spin to update one by one.
However, it is difficult to apply this method for particle systems,
since neighbors of particles are not fixed.
Here we use a complete binary tree search for the choosing 
part of the algorithm.

First, calculate the area $A_i$ for each of the particles.
Since an acceptance probability $1-\lambda_i$ is 
proportional to $A_i$ as shown in Eq.~(\ref{DefLamda_i}),
the particle should be chosen with the probability 
proportional to $A_i$.

Next, construct a complete binary tree as follows,
\begin{enumerate}
\item Prepare a complete binary tree with enough height $h$,
this height $h$ should satisfy $2^{h-2} < N \le 2^{h-1}$.
\item Label each node with $T_n^k$, which denotes 
the $n^{\rm th}$ value at level $k$.
The root node is labeled by $T_1^{h}$.
A node labeled $T_n^{k+1}$
has branches leading to two nodes $T_{2n-1}^k$ and $T_{2n}^k$.
\item Associate every bottom node $T_i^1$ with the value of area $A_i$.
If the number of bottom nodes $2^{h-1}$ is larger than $N$,
the rest of the 
nodes are associated with zero, namely, $T_i^1 = 0 ~(i > N)$.
\item Associate nodes at higher levels ($k>1$) recursively with 
the sum of the values associated with its two children,
namely, $T_n^{k+1} = T_{2n-1}^k+T_{2n}^k$.
\end{enumerate}

\begin{figure}[htbp]
\includegraphics[width=0.9\linewidth]{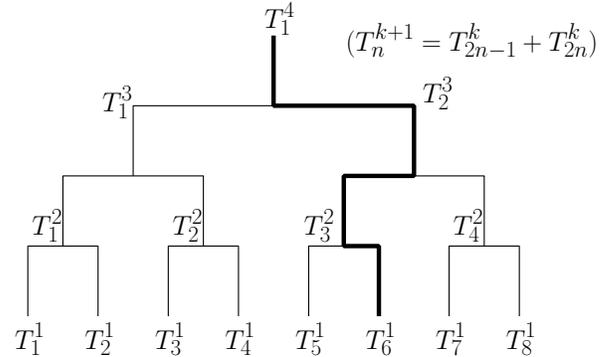}
\caption{
A complete binary tree search. An example of $N=8$ ($h=4$) is shown.
The value of a node $T_i^k$ is the sum of the values of 
its two child nodes, 
namely, $T_i^k=T_{2i}^{k-1}+T_{2i-1}^{k-1}$.
After construction of the tree, we use $h-1$ 
random numbers to choose a bottom 
node.  This bottom node is 
associated with the index of the particle that will be moved 
in this algorithmic step of the RFMC method. 
}
\label{fig_treesearch}
\end{figure}

A sample of a compete binary tree is shown in Fig.~\ref{fig_treesearch}.
Each node has the value $T_n^k$ and the value of each node at level $k+1$
is the sum of the values of its two children nodes at level $k$.
The root node, which is $T_1^4$ in Fig.~\ref{fig_treesearch},
has the sum of all $A_i$, that is,
\begin{equation}
T_1^{h} = \sum_{i}^{N} A_i.
\label{Tree2}
\end{equation}
Using this tree, we can choose a particle with the probability
proportional to $A_i$ in the following way.
\begin{enumerate}
\item $k \leftarrow 1$, $i \leftarrow 1$.
\item Prepare a random number $r$ uniform on $(0,T_i^k)$.  \label{b_tree}
\item $\left\{
\begin{array}{lc}
i \leftarrow 2i-1 & \quad \mbox{if} \quad r<T_{2i-1}^{k-1} \\
i \leftarrow 2i   & \quad \mbox{otherwise}
\end{array}
\right.
$
\item $k \leftarrow i-1$.
\item if $k>1$ then go to $\ref{b_tree}$
\end{enumerate}
Consequently, choosing the bottom node requires $h-1$ random numbers.  

After the above processes, the final value of $i$
indicates the index of the particle to move.
The order of this search algorithm is $O(\log N)$.
When the position of particle $i$ is moved,
the value of $A_i$ is also modified.
We only have to update part of this tree for the chosen particle and 
its neighbors. The order of this update is also  $O(\log N)$,
which is much faster than $O(N)$ of the direct implementation.
Details to implement the complete binary tree search method are
described in the appendix.

\subsection{Find a new position of the particle}

After choosing a particle, we have to choose the new position for it.
It is difficult to choose a point uniformly from 
the points allowed to move into, since its shape is generally
very complicated (see Fig.~\ref{fig_ai}).
Therefore, we have chosen to 
choose the new position using a Monte Carlo rejection 
method. Namely we generate a random position uniformly over some 
bounding area 
that includes all of the area $A_i$. [Such as the 
dashed circle of radius $\sigma_s$ in Fig~\ref{fig_survival}(a) or the 
rectangle in 
Fig~\ref{fig_survival}(c).]  
If this point is not in $A_i$ it is rejected and another 
uniformly distributed random point over the bounding area is generated.  
The first point not be rejected is the new position of the particle, 
since it is in the area $A_i$, 
and this point is used as the new center for the particle. 
This completes one algorithmic step of the RFMC method.  

A value of area $A_i$ is very small compared to 
the trial circle at high density, and hence the 
Monte Carlo trial to find the new position of the particle 
to be moved 
became very inefficient.  
To improve this, it is effective to limit the trial area 
for the Monte Carlo by making the bounding area very close to 
the area $A_i$. 
We outline two different survived point 
methods, but have only implemented the first.

For the first method, the one actually implemented in this paper 
see Fig.~\ref{fig_survival}~(c). The solid rectangle is
a bounding rectangle which includes the area $A_i$.
It is easy to obtain the bounding rectangle with the survived vertex points.  
With the set of survived vertex points $\{(x_i,y_i) \}$, a diagonal 
line of the
bounding rectangle is from $(\min{\{x_i\}},\min{\{y_i\}})$ to 
$(\max{\{x_i\}},\max{\{y_i\}})$.
Then we can perform Monte Carlo trials for a new position 
within only this rectangle. The area of the rectangle is on the 
same order of $A_i$, so the probability of 
success to obtain the new position
is drastically improved compared with the direct search over the 
trial circle.  

An alternative method is first use a random number to decide which of 
the triangles formed with point O and two adjacent survived vertex points the 
survived point will fall into.  This is done analytically since the 
area of each of these triangles with removed shadow circle chords have been 
already calculated.  Then the shortest side formed with point O and the 
two survived vertex points (say $\overline{SP_2}$ in 
Fig.~\ref{fig_survival}~(d)) is lengthened to be equal to the longest side 
($\overline{SP_1}$ in Fig.~\ref{fig_survival}~(d)).  The 
random trial point is then generated within the 
section of the circle with a radius equal to the 
longest side ($\overline{SP_1}$ in Fig.~\ref{fig_survival}~(d)).  
Then the point becomes the survived point used for the new location of 
point O if the trial point is within the shaded area.  Otherwise, this 
procedure repeats in the same extended circular section until 
a survived point is found.  

\section{Results}

\subsection{Calculation of $A_i$}

\begin{figure}[htbp]
\includegraphics[width=0.9\linewidth]{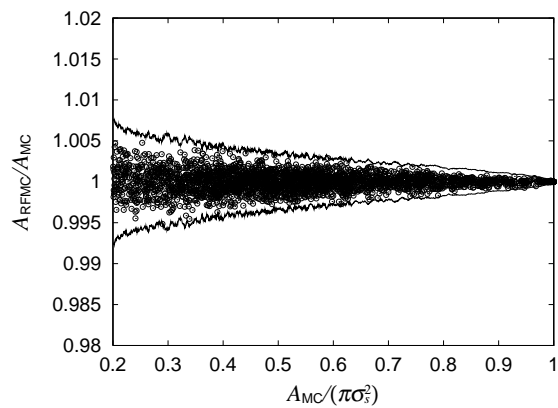}
\caption{Comparison of calculated $A_i$ between our survived points method 
and that calculated 
by a more straightforward MC method.
Units on both axes are dimensionless.
The circles denote the ratio of $A_{\mbox{RFMC}}$ to the $A_{\mbox{MC}}$.
The number of configurations is $3000$ at $\rho = 0.9$ 
and $10^6$ independent samples are averaged for each configuration.
The solid lines are the standard deviation obtained from the jack-knife 
procedure described in the text.
}
\label{fig_afrac}
\end{figure}

In order to test our method to calculate $A_i$ described in 
Sec.~\ref{sec_calcarea},
the values of $A_i$ were also evaluated by a 
Monte Carlo sampling ($A_{\rm MC}$)
with trial points uniformly drawn over the trial circle.  
The density of the system $\rho$ is defined to be
$\rho = 4N\sigma^2 / L^2$
with the number of particles $N$, the radius of the particles $\sigma$ and 
the linear system size $L$, respectively.
Throughout this study, the number of particles $N$ is set to be $23288$
and the periodic boundary condition is taken for both axes.
The number of the generated configurations were 3000,
and $10^6$ MC trial points are taken for each of the 
configurations to evaluate its area.
The density of the system is fixed at $\rho = 0.9$.
The result is shown in Fig.~\ref{fig_afrac}.
The area $A_i$ is normalized by the area of the trial circles 
(see Fig.~\ref{fig_ai}).
The difference between the MC and our survived point method 
is less than $0.01\%$ 
for all areas, which is the same order as the statistical 
error of our MC method.
The standard deviation of the MC area calculation is determined
by dividing the data into 10 groups, each including $10^5$ samples.
This result shows that the value of $A_i$ is properly 
calculated by our method.

\subsection{Time evolution}

\begin{figure}[htbp]
\includegraphics[width=0.9\linewidth]{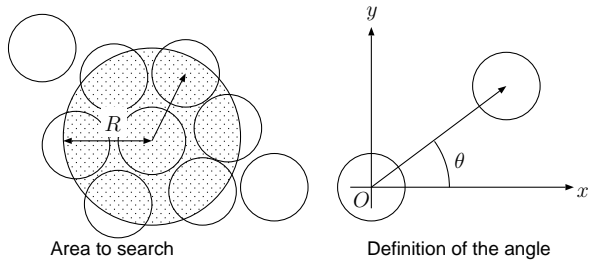}
\caption{
The definition of the neighboring particles and 
bond angle $\theta$.
Two particles separated by a distance less than $R$ 
are defined as neighbors. Here, $R$ is set to be $2.6\sigma$
with the radius $\sigma$ of particles.
The bond angle $\theta$ is defined to be an angle 
between the bond connecting neighboring particles with respect 
to an arbitrary, but
fixed global axis. 
}
\label{fig_neighbours}
\end{figure}

To compare the dynamics between the standard MC and the RFMC,
we observed the time evolution of the six-fold 
bond-orientational order parameter $\phi_6$ \cite{HalperinNelson}.
The parameter $\phi_6$ is defined to be
\begin{equation}
\phi_6 =  \left< \exp(6 i \theta) \right>,
\end{equation}
with the bond angle $\theta$ which has a definition described in
Fig.~\ref{fig_neighbours}. 
The average is taken for all pairs of neighboring particles.
The parameter $\phi_6$ becomes 1 when all particles are 
located on the points of a hexagonal grid, and it becomes $0$ 
when the particle location is completely disordered.
Therefore $\phi_6$ describes
how close the system is to the perfect hexagonal packing.
The neighbors in an off-lattice model are strictly defined with the
Voronoi construction \cite{Jaster}, which is a very time-consuming method.
In this paper, two particles separated by a distance less than $2.6\sigma$ 
are defined as neighbors. We confirmed that the value of $\phi_6$ is
approximately the same value as the value obtained 
with the Voronoi construction.
At the beginning of the simulation, the particles are set up 
in a perfect hexagonal order, namely, $\phi_6(t=0, \rho) = 1$.
The order parameter $\phi_6$ starts to relax to
the value of the equilibrium state.
With this nonequilibrium relaxation (NER) behavior
of order parameters, critical points
and critical exponents of various phase transitions
can be determined accurately 
\cite{ItoPhysica1,ItoPhysica2,ItoMastuhisa,ItoHukushima}.
This method is called a NER method.
Watanabe {\it et al.} \cite{Watanabe,Watanabe2} studied 
two-dimensional melting 
based on the NER method for the Kosterlitz-Thouless transition \cite{NERKT}
by observing the relaxation behavior of $\phi_6$.
Therefore, the following time evolutions of $\phi_6$ contains 
information about the two-dimensional melting transition.

\begin{figure}[htbp]
\includegraphics[width=0.95\linewidth]{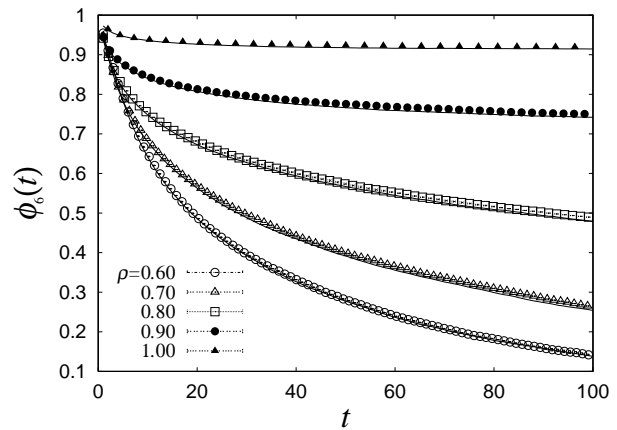}
\caption{Relaxation behaviors of the bond-orientational order.
Solid lines are the results of the standard MC
and symbols (circles, triangles and squares) are the results of the RFMC. 
The data intervals between accepted updates for the RFMC algorithm 
becomes longer at high density, while the data keeps the 
behavior of the standard MC algorithm.
}
\label{fig_density}
\end{figure}

Time evolutions of $\phi_6$ are shown in Fig.~\ref{fig_density}.
Solid lines are results of the standard MC simulation and symbols (circles, 
triangles and squares) 
are the results of the RFMC. Fig.~\ref{fig_density} 
shows that both behaviors are equivalent for the two methods. 
This is essentially a check of the program implementation, since the 
physical dynamic is the same for both the MC and the RFMC methods.  

\subsection{Efficiency}

\begin{figure}[hbt]
\includegraphics[width=0.95\linewidth]{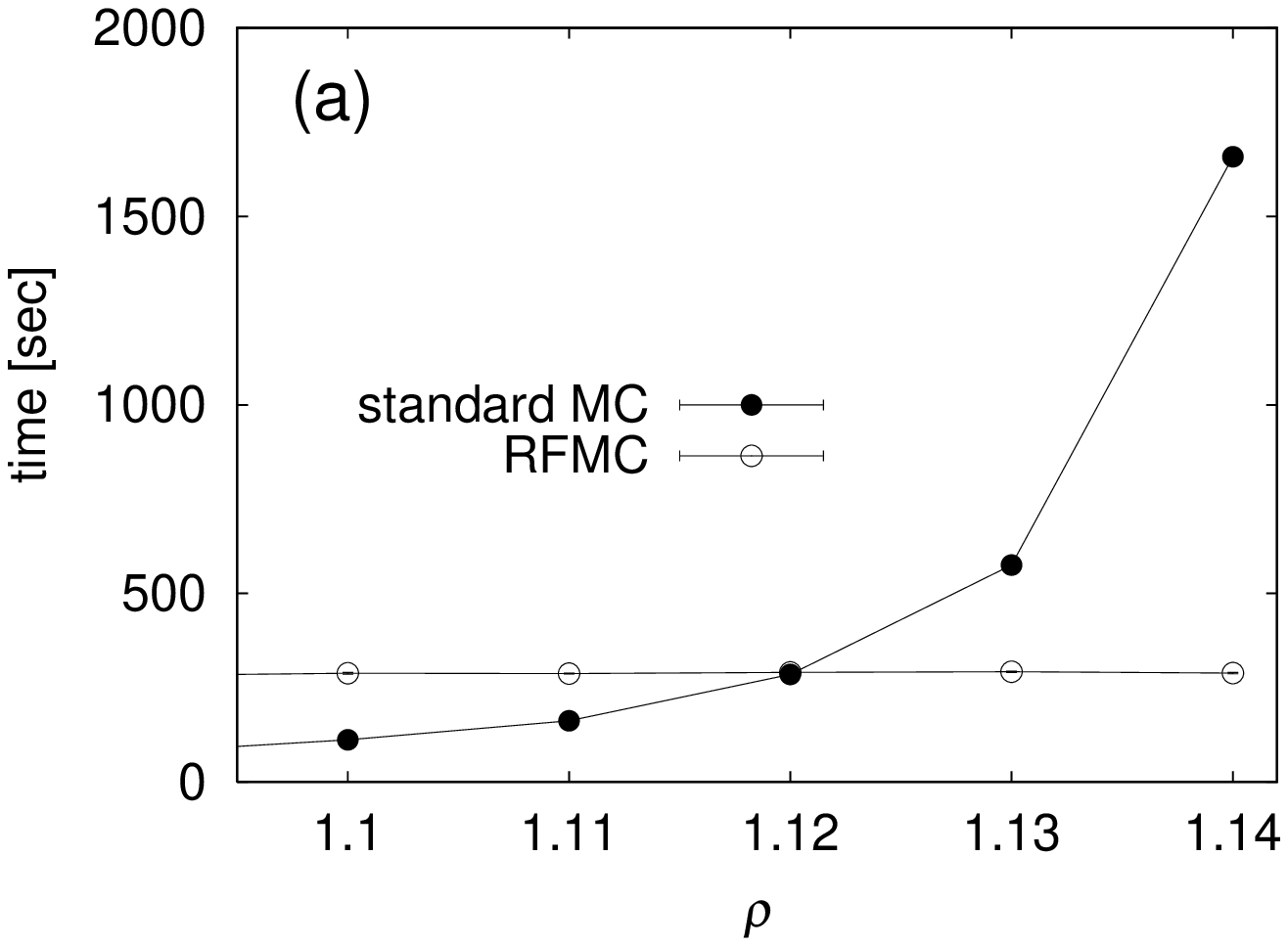}
\includegraphics[width=0.95\linewidth]{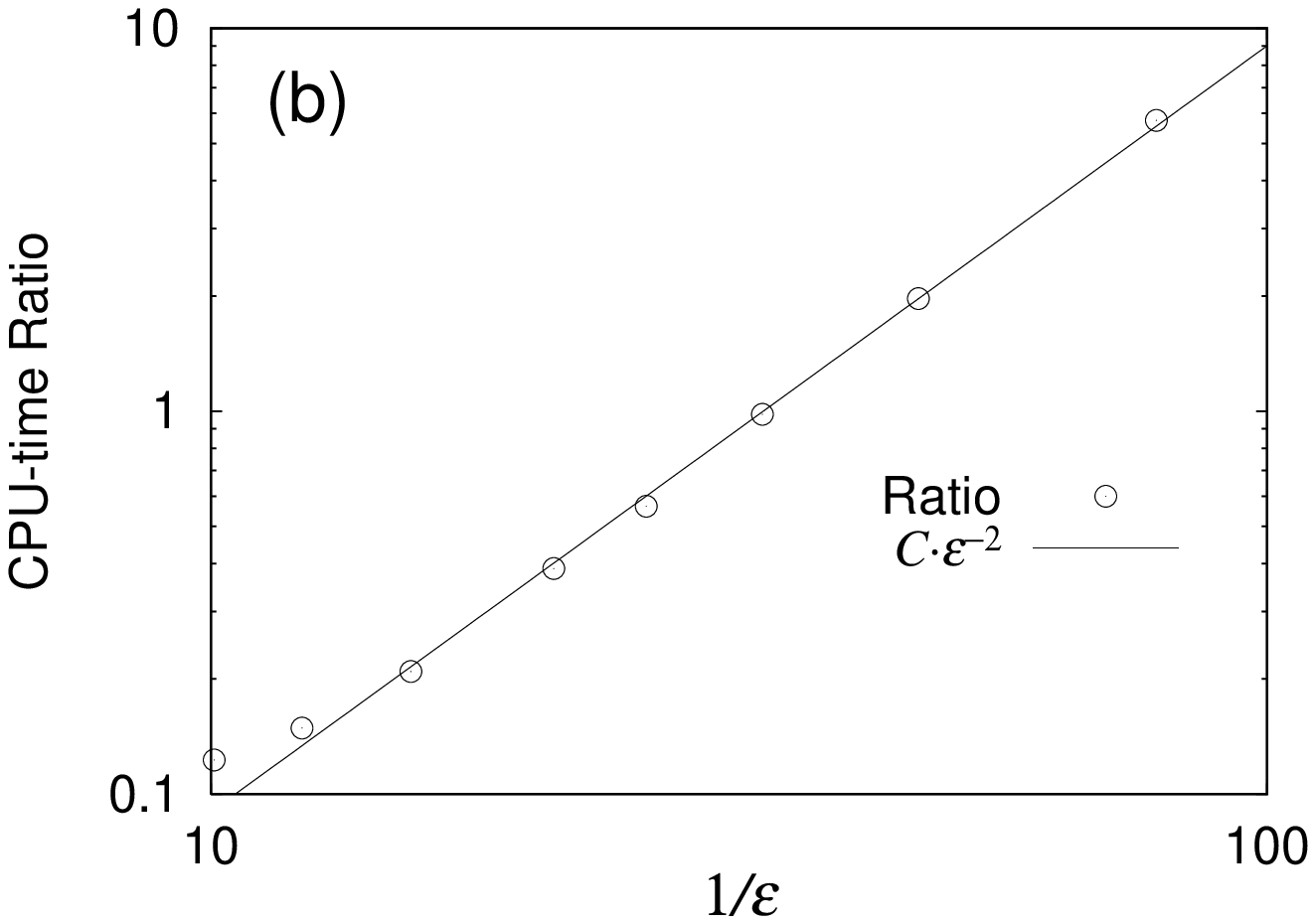}
\caption{
(a) The required computational time to achieve 1000
acceptances of the Monte Carlo moves with the standard 
MC (open circles) and the RFMC (solid circles).
(b) CPU-time ratio vs.~$1/\varepsilon$ with 
$\varepsilon \equiv (\rhoc-\rho)/\rhoc$.
Decimal logarithms are taken for both axes.
The solid line is drawn for the visual reference ($C=0.9 \times 10^{-3}$).
}
\label{fig_speed}
\end{figure}

The computational times required to achieve $1000$ accepted MC steps 
are shown in Fig.~\ref{fig_speed}(a).
Configurations are started from the perfect hexagonal configuration and
both measurements are started after $10^6$ MC steps.
All simulations are performed on an Intel Xeon $2.4$ GHz computer.
While the computational time of the RFMC (open circles) is
almost constant, a longer computational time is required 
for the standard Monte Carlo (solid circles) at higher density. 
It shows that the RFMC is more efficient at high densities,
in spite of the additional bookkeeping involved
in the RFMC method (so one algorithmic step takes 
much longer than one standard MC step).

The CPU-time ratio of the standard MC to the RFMC is shown in 
Fig.~\ref{fig_speed}(b). The data are shown as a function of 
$1/\varepsilon$, where $\varepsilon \equiv (\rhoc-\rho)/\rho_c$, 
the closest density is $\rhoc$ and the density of the 
system is $\rho$.  
The CPU-time ratio, which is the efficiency of the RFMC
compared to that of the standard MC, diverges as $\varepsilon^{-2}$.

\section{Summary and Discussion}

We constructed a rejection-free Monte Carlo algorithm for the 
hard-disk system.
This method conserves the property of the dynamic behavior of 
the original Monte Carlo method.  In other words, the 
time scales will not depend on the density, but are rather set 
by some Brownian-motion type of dynamic for all densities.
An estimate of the time scales between the MC and physical time 
can thus be obtained 
by setting the mean-free path of an isolated particle to be 
proportional to the value $\sigma_s$.  Note that strictly this is only true 
in the limit $\sigma_s\rightarrow 0$, but it should be a reasonable 
approximation for a small finite $\sigma_s$.

We also find that for a fixed value of $\sigma_s$, the RFMC method 
is more efficient at high density. Therefore, the RFMC method 
should be useful for 
studies of two-dimensional solids or studies of high-density glass materials.
It may also be possible to make the algorithm even more efficient by 
utilizing the ideas of absorbing Markov chains (for the MCAMC method 
for discrete state spaces see \cite{Novotny} and references therein).  
Increased algorithmic efficiencies for the Monte Carlo dynamics 
of hard disks could be useful to further test physical phenomena using 
hard disk systems, such as for example 
the relationship between fluctuations and 
dissipation of work in a Joule experiment \cite{Cleuren}.  

\section*{Acknowledgement}

The authors thank S.\ Miyashita, P.A.\ Rikvold, and S.\ Todo 
for fruitful discussions.
This work was partially supported by the Grant-in-Aid for
Scientific Research (C), No.~15607003, of Japan Society for
the Promotion of Science, and the Grant-in-Aid for Young
Scientists (B), No.~14740229, of the Ministry of Education,
Culture, Sports, Science and Technology of Japan,
the 21st  COE program, ``Frontiers of Computational Science", 
Nagoya University
and by the U.S.\ NSF grants DMR-0120310, DMR-0426488, and DMR-0444051.  

\section*{Appendix}

\begin{figure}[tbh]
\includegraphics[width=0.9\linewidth]{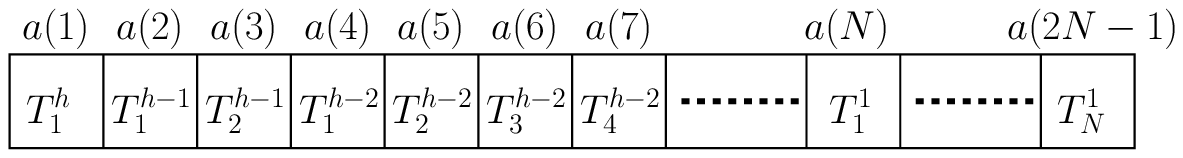}
\caption{
Implementation of the complete binary tree search with an array.
The required size of the array to implement the tree with height $h$
is $2^h-1$. The height $h$ should satisfy $2^{h-2} < N \leq 2^{h-1}$.
When the number of particles $N$ is $2^{h-1}$, which is the 
maximum number of particles that the tree with height $h$ can treat,
the size of the array is $2N-1$.
}
\label{fig_treeimplement}
\end{figure}

The complete binary tree search can be implemented with 
an one-dimensional array. 
To make it simple, let the number of particles $N$ be $2^{h-1}$.
The tree with height $h$ requires an array $a(i)$ with size $2N-1$.
First, associate each bottom node with a corresponding value as
\begin{equation}
a(N+i-1) \leftarrow A_i \quad (i=1,2,\cdots,N),
\end{equation}
which corresponds to $T_i^1 \leftarrow A_i$.
Next, associate parent nodes recursively as 
\begin{list}{}{}
\item $i \leftarrow N$
\item While $i \neq 0$
\item \quad $a(i) \leftarrow a(2i) + a(2i+1)$
\item \quad $i \leftarrow i - 1$
\item next $i$
\end{list}
Using this array, we can pick up particle $i$ with the probability 
proportional to $A_i$ as,
\begin{list}{}{}
\item $i \leftarrow 1$
\item While $i < N$
\item \quad Prepare a uniform random number $r$ on $(0,a(i))$
\item \quad
$
\left\{
\begin{array}{ll}
i \leftarrow 2i & \quad \mbox{if} \quad r < a(2i) \\
i \leftarrow 2i+1   & \quad \mbox{otherwise}
\end{array}
\right.
$
\item next $i$
\item $i \leftarrow i - N +1$.
\end{list}
After the above procedure, we obtain the index $i$ of 
the chosen particle.
When the value of $A_i$ is changed,
the tree should be updated.
The update process is as follows,
\begin{list}{}{}
\item $a(N+i-1) \leftarrow A_i$
\item $i \leftarrow \left\lfloor \displaystyle\frac{i + N}{2} \right\rfloor$
\item While $i \neq 1$
\item \quad $a(i) \leftarrow a(2i)+a(2i+1)$
\item \quad $i \leftarrow \left\lfloor \displaystyle i/2 \right\rfloor$
\item next $i$.
\end{list}
Note that, when the chosen particle is moved,
the acceptance probabilities of 
the neighboring particles of the moved particle are also changed.
Therefore, we have to perform the above process for all neighboring particles.



\begin{thebibliography}{99}
\bibitem{IntegrationSearch}
The order of the search itself is $O(\ln N)$
in the direct implementation with the integration scheme, 
since we can use binary search for an ordered list.
However, when one value is modified, the order of the
computation to update the list of values is $O(N)$.
\end{thebibliography}
\end{document}